\documentclass[preprint,aps]{revtex4-2}
\usepackage{mathrsfs}
\usepackage{graphicx}
\usepackage{multirow}
\usepackage{makecell}
\usepackage{booktabs}
\usepackage{hyperref}
\hypersetup{
    colorlinks,
    citecolor=black,
    filecolor=black,
    linkcolor=black,
    urlcolor=black
}
\usepackage{color}
\usepackage{bm}

\begin{document}

\title{Extreme Strain Controlled Correlated Metal-Insulator Transition in the Altermagnet CrSb}

\author{Cong Li$^{1,\sharp}$, Mengli Hu$^{2}$, Jianfeng Zhang$^{3}$, Magnus H. Berntsen$^{1}$, Francesco Scali$^{4}$, Dibya Phuyal$^{1}$, Chun Lin$^{5,6}$, Wanyu Chen$^{1}$, Johan Chang$^{5}$, Oliver J. Clark$^{7}$, Timur K. Kim$^{7}$, Jacek Osiecki$^{8}$, Craig Polley$^{8}$, Balasubramanian Thiagarajan$^{8}$, Zhilin Li$^{9,\sharp}$, Tao Xiang$^{9,\sharp}$ Oscar Tjernberg$^{1,\sharp}$
}

\affiliation{
\\$^{1}$Department of Applied Physics, KTH Royal Institute of Technology, Stockholm 11419, Sweden
\\$^{2}$Leibniz Institute for Solid State and Materials Research, IFW Dresden, Helmholtzstraße 20, 01069 Dresden, Germany
\\$^{3}$Center for High Pressure Science and Technology Advanced Research, Beijing 100193, China.
\\$^{4}$Dipartimento di Fisica, Politecnico di Milano, Piazza Leonardo da Vinci 32, 20133 Milano, Italy
\\$^{5}$Physik-Institut, Universit\"{a}t Z\"{u}rich, Winterthurerstrasse 190, Z\"{u}rich CH-8057, Switzerland
\\$^{6}$Stanford Synchrotron Radiation Lightsource, SLAC National Accelerator Laboratory, Menlo Park, California 94025, USA
\\$^{7}$Diamond Light Source, Harwell Campus, Didcot, OX11 0DE, United Kingdom
\\$^{8}$MAX IV Laboratory, Lund University, 22100 Lund, Sweden
\\$^{9}$Beijing National Laboratory for Condensed Matter Physics, Institute of Physics, Chinese Academy of Sciences, Beijing 100190, China
\\$^{\sharp}$Corresponding authors: conli@kth.se, lizhilin@iphy.ac.cn, txiang@iphy.ac.cn, oscar@kth.se
}


\maketitle


\begin{center}
{\bf Abstract}
\end{center}

{\bf Correlated flat bands and altermagnetism are two important directions in quantum materials, centred respectively on interaction-dominated phases and symmetry-enforced spin-textured states, yet both derive from lattice symmetry and orbital hybridization. This common origin implies that extreme crystal distortion, by narrowing bandwidths, enhancing correlations and reshaping the symmetries of altermagnetic spin splittings, could unify flat-band and altermagnetic physics in a single material; in practice, however, achieving such large distortions in a crystalline altermagnet is a formidable challenge. Here we combine a dedicated strain device with a tailored single-crystal mounting scheme to impose a highly tensile strain gradient in bulk CrSb, a prototypical altermagnet, creating a near-surface layer in which the in-plane lattice is strongly distorted relative to the weakly strained bulk, while the average bulk distortion remains small. Angle-resolved photoemission reveals a reversible regime at moderate strain, where a deeper flat-band feature, attributed to a strain-gradient-driven suppression of Cr-Sb hybridization, coexists with a correlation-enhanced Cr 3$d$ flat band, and an irreversible regime at larger strain where partial bond decoupling drives a predominantly insulating spectral response. Density-functional calculations show that an orbital-selective altermagnetic spin texture persists across this correlated regime despite strong bandwidth renormalisation. These results define a strain-symmetry-correlation map for CrSb and establish extreme tensile strain as a route to co-engineer flat-band tendencies and spin-textured, zero-net-moment correlated states in altermagnets, pointing toward strain-adaptive, spin-selective Mott filtering and related device concepts.\\
}


{\bf Introduction}

Lattice geometry and symmetry have become primary design variables for organising electronic correlations and spin textures in quantum materials. Correlated electronic flat bands in moir\'e and kagome systems\cite{YCao_Nature2018_1_PJHerrero,YCao_Nature2018_2_PJHerrero,MGKang_NM2020_RComin,YHu_NC2022_MShi,JPWakefield_Nature2023_JGCheckelsky,JGCheckelsky_NRM2024_SPaschen,CChen_Nature2024_YLChen,JWHuang_NP2024_MYi,QLi_NM2024_SYZhou,LChen_NC2024_QMSi} illustrate how suppressing kinetic energy magnifies Coulomb and Hund interactions and produces bandwidth-controlled instabilities, so that modest changes in bond geometry or stacking can switch between superconducting, magnetic and topological ground states. Altermagnets\cite{LSmejkal_PRX2022_1_TJungwirth,LSmejkal_PRX2022_2_TJungwirth,JKrempasky_Nature2024_TJungwirth,SLee_PRL2024_CYKim,TOsumi_PRB2024_TSato,OJAmin_Nature2024_PWadley,SReimers_NC2024_MJourdan,OFedchenko_SA2024_HJElmers,HReichlova_NC2024_LSmejkal,ZYZhou_Nature2025_CSong,BJiang_NP2025_TQian,FYZhang_Nature2025_CYChen,CLi_CP2025_JCDBrink,CSong_NRM2025_FPan,WTLin_AM2025_LMiao} provide a complementary route, realising collinear magnetic order with zero net magnetisation in which crystalline symmetry enforces spin splitting and an orbital-selective spin texture in the bands\cite{LSmejkal_PRX2022_1_TJungwirth,LSmejkal_PRX2022_2_TJungwirth}. In both cases, lattice symmetry constrains orbital hybridization and correlation, suggesting that bandwidth engineering and spin-texture engineering share a common microscopic basis rather than representing independent problems.

Within this structural-control viewpoint, lattice distortion is a particularly powerful and conceptually unifying control knob. On the flat-band side, pressure, uniaxial strain and controlled heterostrain tune the effective bandwidth-to-interaction ratio, move van Hove singularities through the Fermi level and reorder orbital hierarchies\cite{JPWakefield_Nature2023_JGCheckelsky,JGCheckelsky_NRM2024_SPaschen,MGKang_NM2020_RComin,YHu_NC2022_MShi,JWHuang_NP2024_MYi,ZBi_PRB2019_LFu,HLaBollita_PRB2021_ASBotana,AConsiglio_PRB2022_DDiSante,CLin_NC2024_JChang}, thereby stabilising or suppressing correlated insulators, unconventional superconductors and topological states\cite{YCao_Nature2018_1_PJHerrero,YCao_Nature2018_2_PJHerrero,CChen_Nature2024_YLChen,QLi_NM2024_SYZhou,LChen_NC2024_QMSi,LZhang_NL2022_JTYe,QWang_AM2021_YPQi}. For altermagnets, theory shows that spin splitting, Berry curvature and spin-orbit responses are dictated by symmetry-allowed hybridization channels\cite{LSmejkal_PRX2022_1_TJungwirth,LSmejkal_PRX2022_2_TJungwirth,CLi_CP2025_JCDBrink,BKaretta_PRB2025_JSinova,LBai_AFM2024_YGYao,PAMcClarty_PRL2024_JGRau,PGRadaelli_PRB2024} and are consistent with a rich spectrum of spin-orbit and symmetry-enabled transport and spectroscopic phenomena already reported in this class of materials\cite{HMa_NC2021_JLiu,RGHernandez_PRL2021_JZelezny,SKarube_PRL2022_JNitta,HBai_PRL2022_CSong,HBai_PRL2023_CSong,LHan_SA2024_FPan,ALeon_npj2025_JWGonzalez}. From this common microscopic perspective, extreme crystal deformation is especially appealing: a single structural field can narrow bandwidths, enhance correlations and reshape the symmetry constraints that control the altermagnetic spin splitting, raising the prospect of co-tuning flat-band tendencies and altermagnetic spin textures\cite{HMa_NC2021_JLiu,RGHernandez_PRL2021_JZelezny,SKarube_PRL2022_JNitta,HBai_PRL2022_CSong,HBai_PRL2023_CSong,LHan_SA2024_FPan,ALeon_npj2025_JWGonzalez,KDBelashchenko_PRL2025,AChakraborty_PRB2024_JSinova,JDSForte_Arxiv2025_TLow,KTakahashi_PRB2025_JSchmalian,SGJeong_Arxiv2025_BJalan,JYSun_Arxiv2025_EJKan} within one material platform.

In practice, however, the accessible strain window in crystalline altermagnets has been very limited. Conventional mechanical loading of bulk crystals typically reaches only modest strains before fracture, while lattice control of altermagnets has so far been realised mainly in epitaxial thin films and heterostructures, where lattice mismatch and interfacial strain are typically limited to $\sim 1\%$\cite{ZYZhou_Nature2025_CSong,WTLin_AM2025_LMiao,SReimers_NC2024_MJourdan,SSanthosh_AM2025_NSamarth,SAota_PRM2025_MTanaka,SPBommanaboyena_PRM2025_DKriegner,ABadura_NC2025_HReichlova,SGJeong_Arxiv2025_BJalan,JDSForte_Arxiv2025_TLow}. Such weak strain is far from the regime in which hybridization collapse, bandwidth renormalisation, symmetry breaking and electronic correlations become deeply intertwined and flat or quasi-flat bands can emerge from dispersive manifolds. To overcome this limitation, we design a dedicated strain device\cite{CLin_NC2024_JChang,CLin_NM2021_TKondo} and, together with a tailored single-crystal mounting scheme, concentrate tensile stress into a narrow near-surface region of a bulk crystal, creating a tunable strain-gradient layer in which the in-plane lattice is strongly distorted relative to the bulk while the average bulk distortion remains small. The strain in this layer is reversible and continuously controllable, thereby opening, within a single crystal, the strain window in which correlated flat-band behaviour and symmetry reshaping are expected to become intertwined.

\begin{figure*}[tbp]
\begin{center}
\includegraphics[width=1\columnwidth,angle=0]{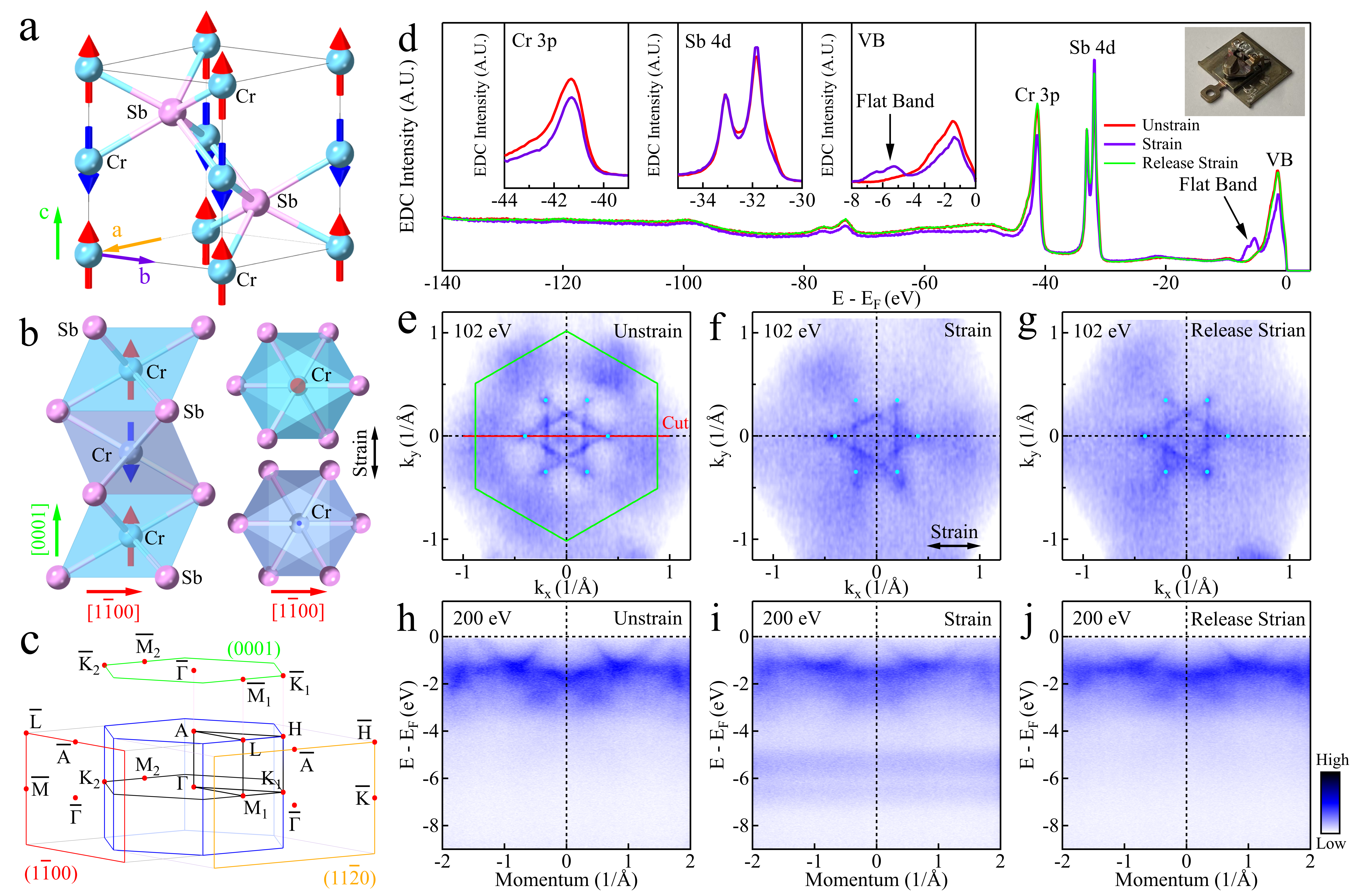}
\end{center}
\caption{\footnotesize\textbf{Crystal structure of CrSb and strain induced flat bands on the (0001) plane.} (a) The crystal structure of CrSb with the space group $P6_3/mmc$ (no. 194). (b) The side and top views of the opposite-spin sublattices. (c) The 3D BZ of the original unit cell of CrSb, and the corresponding two-dimensional Brillouin zone (BZ) projected on the (0001) plane (green lines), (1$\overline{1}$00) plane (red lines) and (11$\overline{2}$0) plane (orange lines). (d) The XPS measurements of CrSb were recorded with a photon energy of 200~eV under applied strain (purple line), without strain (red line), and after strain release (green line). All measurements were normalized over the range -140~eV to -110~eV. The inset presents an enlarged view of the Si 3p orbital, Sb 4d orbital and value band (VB). The inset in the upper right corner shows the sample holder with strain functionality. The black arrows mark the flat band. (e-g) Fermi surfaces of CrSb measured at a photon energy of 102~eV: (e) without applied strain, (f) under applied strain along [11$\overline{2}$0] ($\Gamma$-M) direction, and (g) after strain release. Cyan dots indicate the six vertices of the unstrained Star-of-David Fermi surface. (h-j) Corresponding band structures of CrSb measured at a photon energy of 200~eV along the $\Gamma$-M directon [the red line in (e)]: (h) without strain, (i) with applied strain along [11$\overline{2}$0] ($\Gamma$-M) direction, and (j) after strain release.
}
\label{1}
\end{figure*}

CrSb is an archetypal collinear altermagnet (Fig.~\ref{1}a-\ref{1}b) with well-established symmetry-enforced $g$-wave spin splitting and a relatively simple crystal and electronic structure\cite{LSmejkal_PRX2022_1_TJungwirth,SReimers_NC2024_MJourdan,CLi_CP2025_JCDBrink,WTLin_AM2025_LMiao,BKaretta_PRB2025_JSinova,JXWu_APL2023_XHShao,MZeng_AS2024_CLiu,JYDing_PRL2024_JGCheng,GWYang_NC2025_YLiu,WLLu_NL2025_JZMa,SSanthosh_AM2025_NSamarth,TUrata_PRM2024_HIkuta}, in which the same Cr-Sb and Cr-Cr hybridization paths set both the bandwidth and the symmetry of the spin splitting. This makes CrSb an ideal testbed for our extreme-strain scheme. By applying the near-surface strain-gradient protocol to bulk CrSb, we drive the system into a regime where lattice symmetry, hybridization and electron correlation are strongly entangled, yet the underlying altermagnetic order remains intact. As we show below, combining angle-resolved photoemission spectroscopy (ARPES) with structural probes and density functional theory (DFT) calculations allows us to map out a strain-symmetry-correlation phase diagram, in which emergent flat-band tendencies, correlation-enhanced insulating behaviour and a robust altermagnetic spin texture can be viewed within a single, unified framework. We now turn to the experimental realisation of the strain geometry and its impact on the electronic structure.\\

{\bf Results}


To examine how strain reshapes the electronic structure of altermagnet CrSb, we compare ARPES and X-ray photoelectron spectroscopy (XPS) measurements acquired in the unstrained, strained, and strain-released states. The detailed sample preparation procedure is described in Section~2 of the Supplementary Materials. XPS spectra (Fig.~\ref{1}d) show that the core levels (Cr 3p, Sb 4d) remain nearly unchanged under strain, whereas the valence band (VB) develops two prominent peaks near -6 eV and exhibits a transfer of spectral weight from the Fermi level to deeper binding energies, signaling a strain-driven electronic reconstruction. 

The Fermi surfaces of the (0001) surface, measured at 102 eV photon energy, are shown in Fig.~\ref{1}e-\ref{1}g (see Sections~3 and~4 of the Supplementary Materials for details on the strain-dependent electronic structure evolution). In the unstrained state (Fig.~\ref{1}e), the Fermi surface retains its hexagonal symmetry. Under tensile strain (Fig.~\ref{1}f), it becomes strongly distorted due to the strain. Removing the strain restores the original Fermi surface geometry (Fig.~\ref{1}g), showing that the momentum-space distortion is fully reversible. Complementary band dispersion measured at 200 eV photon energy (Fig.~\ref{1}h-\ref{1}j) reveal the associated high-binding-energy spectral evolution. In the strained state (Fig.~\ref{1}i), two dispersionless features appear near -6~eV, consistent with the peaks observed in XPS and indicating a strain-induced reconstruction of deeply bound states. These flat-band features are absent in both the unstrained (Fig.~\ref{1}h) and strain-released (Fig.~\ref{1}j) spectra, demonstrating their reversibility and indicating that the -6~eV flat-band features are tied to the applied tensile strain rather than to irreversible chemical or structural modification. More detailed photon energy dependent electronic structure measurements of the strained (0001) surface are provided in Section~5 of the Supplementary Materials, while the microscopic origin of the -6~eV flat band will be discussed in detail later.

\begin{figure*}[tbp]
\begin{center}
\includegraphics[width=1\columnwidth,angle=0]{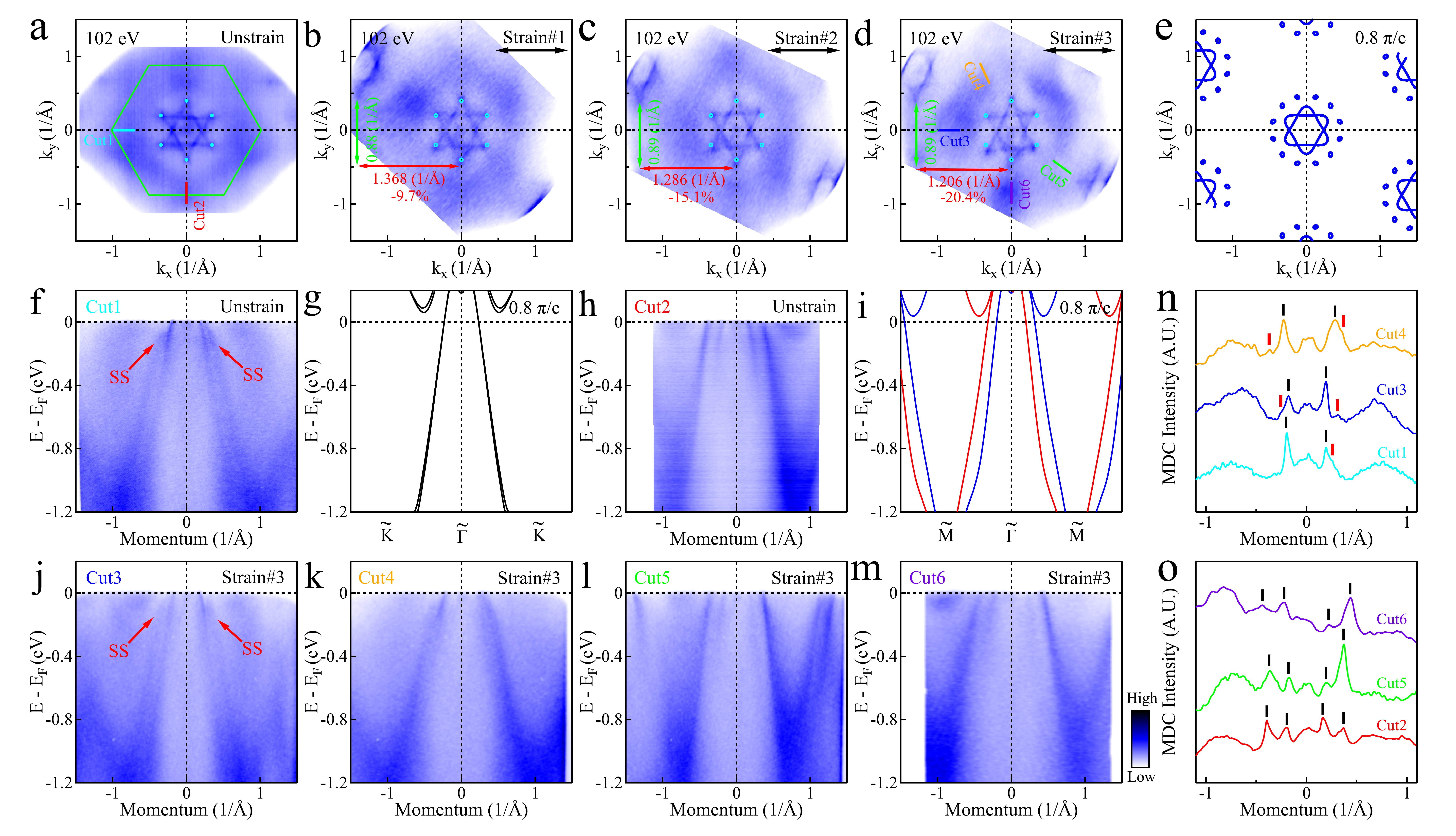}
\end{center}
\caption{\footnotesize\textbf{Strain-induced evolution of the electronic structure near the Fermi level.} (a-d) Fermi surface evolution under applied strain measured with photon energy of 102~eV: (a) shows the Fermi surface without strain; (b–d) correspond to increasing levels of applied strain along the [1$\overline{1}$00] ($\Gamma$-K) directions. Cyan solid and open dots indicate the six vertices of the unstrained Star-of-David Fermi surface. The red and green double-headed arrows indicate the momentum-space distances along k$_x$ and k$_y$ between equivalent Fermi-surface positions in the first and second BZs. (e) The DFT calculated Fermi surface at k$_z$ = 0.8 $\pi$/c. (f, h) Band structures along Cut1 [cyan line in (a), f] and Cut2 [red line in (a), h], measured without strain with photon energy of 102~eV. (g, i) The density functional theory (DFT) calculated band structures along the $\widetilde{K}$-$\widetilde{\Gamma}$-$\widetilde{K}$ (g) and $\widetilde{M}$-$\widetilde{\Gamma}$-$\widetilde{M}$ (i) directions at k$_z$ = 0.8 $\pi$/c. (j-m) Band structures measured with photon energy of 102~eV along Cut3-Cut6 [blue, orange, green and purple lines in (d)] after the third applied strain. The red arrows marks the surface states (SSs). (n-o) The extracted MDCs from band cut1-cut6 at Fermi level. Black ticks indicate the bulk band k$_F$ positions and red ticks indicate the SSs.
}
\label{2}
\end{figure*}

After establishing the strain-induced flat-band formation at deep binding energies, we now turn to the low-energy electronic states near the Fermi level on the (0001) surface (Fig.~\ref{2}). As the tensile strain along $\Gamma$-K is increased, the Fermi surface evolves from a nearly hexagonal contour into an elongated, symmetry-broken shape (Fig.~\ref{2}a-\ref{2}d). The Brillouin zone (BZ) contraction along the strain axis is nearly 20\% while the perpendicular direction remains almost unchanged (Fig.~\ref{2}d). This 20\% value can be viewed as an effective deformation parameter within a simple homogeneous-strain approximation, which we use to quantify the extreme k-space distortion. The associated band dispersions (Fig.~\ref{2}f-\ref{2}m) show that, even under such extreme effective deformation, sharp quasiparticle peaks persist and the overall band periodicity is maintained, consistent with phonon-spectrum calculations (Supplementary Section 9) that confirm the dynamical stability of the lattice under nominal in-plane tensile strains as large as 20\% within our DFT modeling. These observations demonstrate that, within the photoemission probing depth, the in-plane electronic structure experiences a large and highly anisotropic deformation.

To determine whether this deformation permeates the bulk or is confined near the surface, we combine ARPES with X-ray diffraction (XRD). XRD measurements (Supplementary Section~7) show only a 0.5-1\% splitting of the (0002) and (11$\overline{2}$3) reflections, consistent with two bulk structural domains whose lattice parameters differ by at most $\sim$1\% and ruling out a homogeneously strained bulk with larger distortion. The much smaller bulk-averaged distortion compared with the ARPES-inferred effective deformation demonstrates that the large strain seen in ARPES is confined to a near-surface strain-gradient layer bonded to a relatively rigid bulk substrate. Within this gradient, the basal (0001) plane is particularly susceptible to in-plane tensile deformation, which breaks rotational symmetry, perturbs the Cr-Sb coordination and partially suppresses $p$-$d$ hybridization, in a manner consistent with the emergence of the -6~eV flat band. Spatially resolved ARPES further shows that this feature appears only in localized regions and changes markedly upon re-cleaving the same strained crystal (Supplementary Section~7), reinforcing its assignment to a strongly strained near-surface layer. We have reproduced the strain-induced Fermi-surface distortion and the emergence of the -6~eV and -2~eV flat bands in multiple crystals, at two different ARPES beamlines and for both nominal $\Gamma$-M and $\Gamma$-K loading geometries (Supplementary Sections~3-5), demonstrating that these phenomena are robust and reproducible rather than sample- or setup-specific.

Within this extreme-strain state, strain also lifts the momentum-space compensation of the $g$-wave altermagnetic spin texture, producing a strongly anisotropic spin splitting that depends sensitively on crystallographic direction, as captured by spin-resolved DFT (Supplementary Section~8). In the unstrained crystal, opposite-momentum domains yield nearly compensated spin polarisations, suppressing any net spin-splitting effect\cite{BKaretta_PRB2025_JSinova}. Tensile strain removes this compensation and generates a pronounced momentum-dependent spin splitting. This anisotropic altermagnetic state is expected to give rise to directional transport responses\cite{BKaretta_PRB2025_JSinova}, including anisotropic resistivity, mobility and spin-polarised currents, thereby establishing strain as an effective knob for manipulating the altermagnetic spin-splitting effect in CrSb.

\begin{figure*}[tbp]
\begin{center}
\includegraphics[width=1\columnwidth,angle=0]{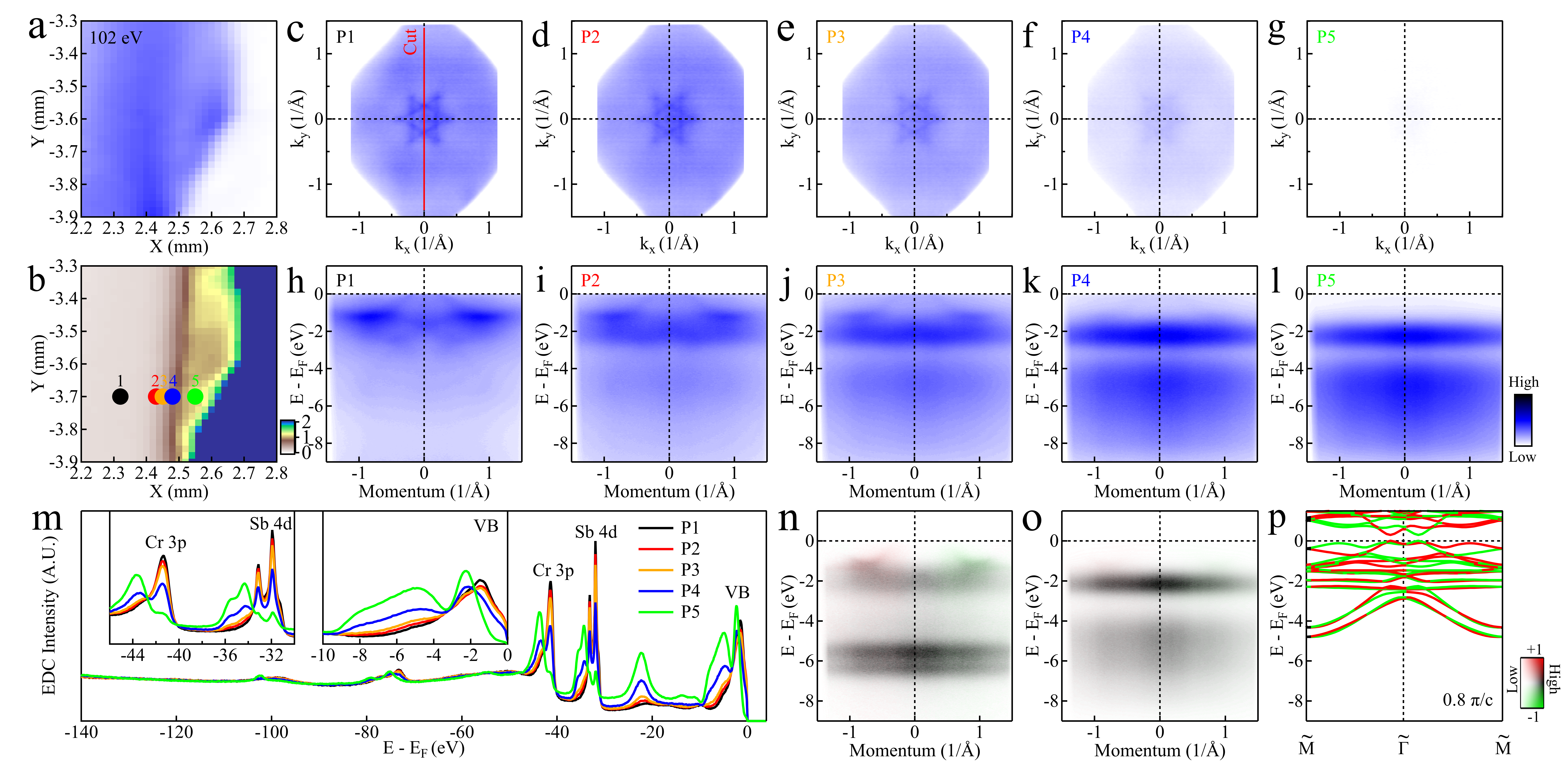}
\end{center}
\caption{\footnotesize\textbf{Spatial evolution of strain-induced flat bands and irreversible spectral response.} (a) Spatial map of the integrated spectral intensity within $\pm$0.75~eV around -1~eV, acquired across a near-surface region with microcracks that delineate a strong strain gradient. (b) Ratio map of the integrated spectral intensity within $\pm$0.75~eV around -6~eV and -1~eV, defining measurement points P1-P5 along the strian gradient. The colored dots mark the beam positions; their sizes approximately represent the beam-spot diameter. (c-g) Fermi-surface mappings at P1-P5 measured with photon energy of 102~eV. (h-l) The corresponding band dispersions along the $\Gamma$-K direction at P1-P5 measured with photon energy of 102~eV. As the measurement position moves from P1 to P5, the spectral weight near the Fermi level gradually decreases, and a distinct flat band emerges around -2~eV, accompanied by the development of a broadened feature near -4.5~eV. Notably, a residual Fermi-surface signal remains visible even where the -2~eV flat band dominates, which likely arises from the photoemission contribution of the beam spot partially overlapping with less-strained metallic regions. (m) XPS spectra measured at P1-P5. The inset presents an enlarged view of the Si 3p orbital, Sb 4d orbital and VB. (n-o) CD-ARPES, plotted as ($I_{CR}$-$I_{CL}$)/($I_{CR}$+$I_{CL}$), for a moderately strained region in the reversible flat-band regime (n) and for a highly strained region in the irreversible regime (o); the irreversible regime at the most strongly strained positions (e.g. P5) likely reflects both defects introduced when cleaving the pre-loaded crystal and additional local damage from the highly non-uniform near-surface stress during tensile strain. (p) Spin-resolved DFT band structure along $\widetilde{M}$-$\widetilde{\Gamma}$-$\widetilde{M}$ under 20\% tensile strain at $k_{z}$ = 0.8 $\pi/c$, with colors indicating opposite $S_z$ projections.
}
\label{3}
\end{figure*}

The analysis above shows that the near-surface strain-gradient layer hosts two characteristic regimes: under sufficiently strong in-plane distortion a pair of flat bands appear near -6~eV, while further increasing the tensile strain brings in an additional flat band around -2~eV accompanied by a gradual depletion of spectral weight at the Fermi level (Supplementary Fig.~S8). To access even larger local strains, we exploit the fact that strong tensile strain produces microcracks on the cleaved (0001) surface, which concentrate stress and enhance the near-surface strain locally. Fig.~\ref{3} shows ARPES measurements along a representative line cut that crosses such a surface microcrack. Along this line, points on one side of the crack probe a weakly strained terrace, while points across the crack sample a strongly strained region that extends beyond the strain window in which the -6~eV structural flat band is the dominant signature (Fig.~\ref{3}a-\ref{3}b). As one moves from the weakly strained side to the strongly strained side, the spectral weight at the Fermi level is progressively depleted (Fig.~\ref{3}c-\ref{3}g), a nearly dispersionless band around -2~eV grows in intensity, and in the strongly strained region a broader flat feature emerges near -4.5~eV (Fig.~\ref{3}h-\ref{3}l). This evolution reflects a strain-driven crossover from an altermagnetic metal towards a regime in which the low-energy response of the near-surface region is dominated by flat bands and localized states.

The XPS taken at the same positions (Fig.~\ref{3}m) shows gradual shifts and, on the strongly strained side, a clear splitting of the Cr~3$p$ and Sb~4$d$ peaks, signalling the emergence of new local chemical environments for Cr and Sb that accompany the electronic reconstruction. The appearance of the -4.5~eV band correlates with these extra core-level components and is naturally interpreted as arising from unbonded Sb~5$p$-derived states created when a subset of Cr-Sb bonds is broken. This assignment is supported by measurements on as-grown, uncleaved CrSb surfaces, where a similar -4.5~eV feature is observed and can be attributed to the 5$p$ orbital of unbonded excess surface Sb atoms (Supplementary Section~7). In the moderately strained, still reversible regime, releasing the external stress restores the original metallic spectrum and removes the -2~eV and -6~eV features (Supplementary Section~4). Under higher strain, by contrast, the -2~eV and -4.5~eV features largely persist after release and the metallic character returns only partially (Supplementary Section~5), indicating that the near-surface region has entered a largely irreversible, insulating-like spectral regime associated with additional structural reconfiguration.

The evolution of the flat bands is mirrored in the dichroic, spin-texture and real-space structural responses. In the reversible regime (Fig.~\ref{3}n), the -2~eV flat band exhibits a strong, momentum-dependent dichroic signal, indicating a well-defined orbital and spin-orbital texture, whereas in the irreversible regime (Fig.~\ref{3}o) this dichroism is strongly suppressed, reflecting the loss of a coherent spin-textured band structure (Supplementary Section~6). Spin-resolved DFT calculations (Fig.~\ref{3}p and Supplementary Section~8) likewise reveal an orbital-selective spin texture associated with the strain-modified Hubbard-band manifold, linking strain gradients, emergent flat bands and the eventual breakdown of coherent altermagnetic order. Consistently, SEM measurements show that strained, cleaved samples develop wrinkles and dense stripe-like textures precisely in regions where the insulating spectral regime is observed, whereas unstrained surfaces remain atomically flat (Supplementary Section~7). These stripe domains are consistent with a martensitic-like transformation\cite{FXiao_SR2018_TFukuda,HMJi_SA2020_PFNealey,YXSong_Nature2025_RKainuma} triggered by excessive tensile deformation, in which local Cr-Sb and Cr-Cr bonds break and long-range lattice coherence collapses.


To clarify the microscopic origin of the reversible flat bands near -2 and -6~eV, we begin by examining the -2~eV feature. DFT calculations with 20\% tensile strain (Supplementary Section~8) reproduce nearly dispersionless Cr~3$d$-derived bands at approximately $\pm 2$~eV, whose energy scale and orbital character closely match the -2~eV flat feature observed in ARPES, supporting its identification as a Mott-like lower Hubbard band (LHB) in a strongly strained altermagnetic lattice. By contrast, Fermi surfaces computed from the same DFT-based Wannier Hamiltonians under uniform 20\% strain show a stronger anisotropic distortion than in experiment (Supplementary Fig.~S28-S29), reflecting the simplified assumption of a homogeneous, fully coherent bulk strain, whereas ARPES probes a near-surface region with strong strain gradients, partial strain relaxation and correlation-induced spectral-weight transfer. Consistently, when only the deep structural flat band near -6~eV is present (Fig.~\ref{1}i), the Fermi-level spectral weight remains largely intact, whereas the emergence and strengthening of the -2~eV flat band are accompanied by a strong suppression of low-energy spectral weight (Fig.~\ref{3} and Supplementary Fig.~S8), in line with correlation-enhanced localization of Cr~3$d$ states in the highly strained near-surface layer.

\begin{figure*}[tbp]
\begin{center}
\includegraphics[width=1\columnwidth,angle=0]{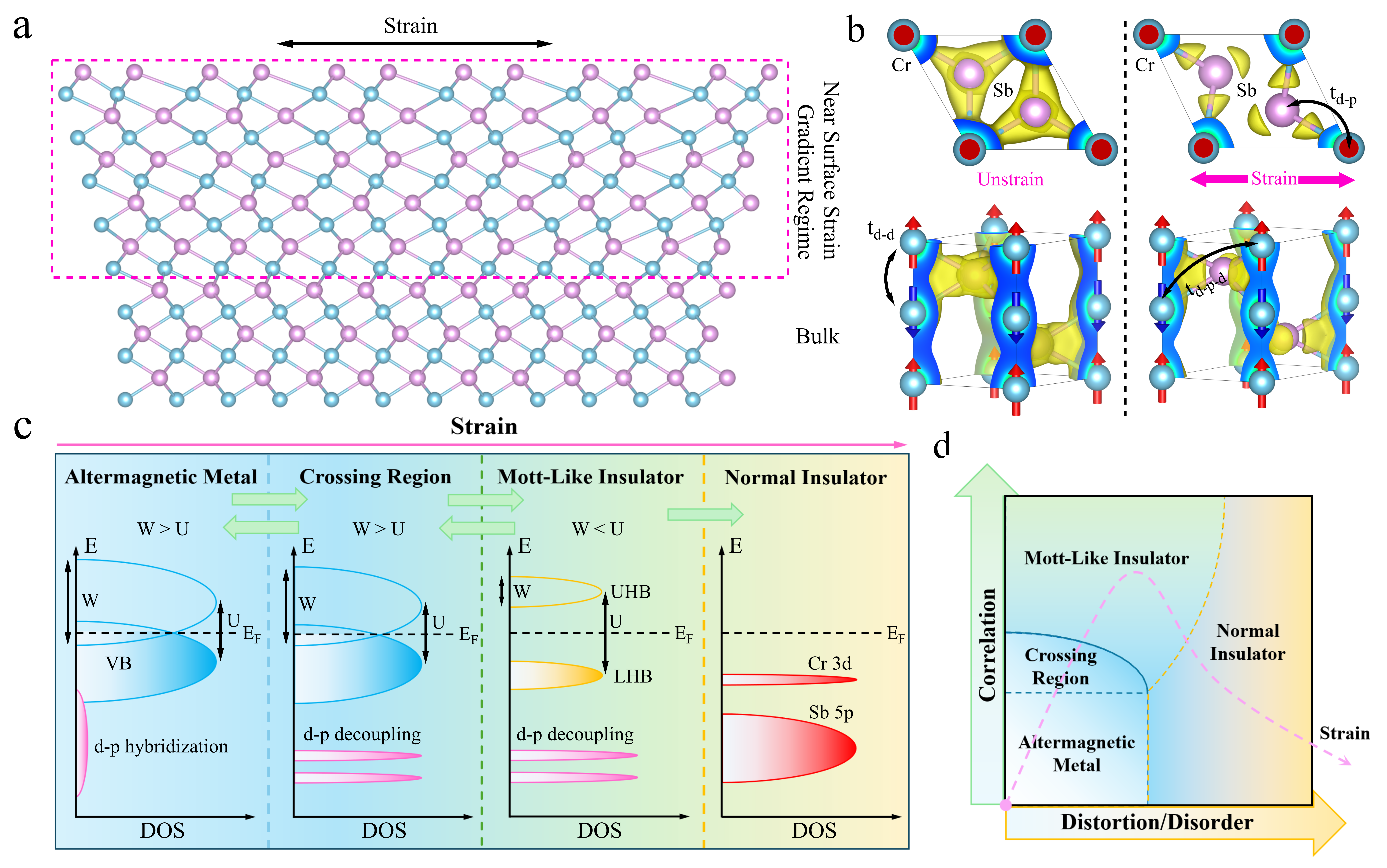}
\end{center}
\caption{\footnotesize\textbf{Strain-controlled evolution of electronic states and correlated phases in CrSb.} (a) Illustration of the near-surface strain-gradient geometry: direct tensile deformation mainly stretches the top several Cr–Sb layers, while the underlying bulk remains weakly distorted. (b) Real-space charge density distributions for the unstrained (left) and 20\% tensile-strained (right) lattices. The isosurfaces, primarily derived from Cr~3$d$ and Sb~5$p$ states from -6 eV to -2 eV, illustrate that tensile strain elongates Cr-Sb bonds and weakens $p$-$d$ hybridization while modulating Cr-Cr $d$-$d$ coupling, leading to enhanced orbital localization. (c) Schematic illustration of the strain-dependent evolution of the electronic density of states (DOS). With increasing tensile strain, CrSb evolves from an altermagnetic metal with dispersive bands to a distorted metallic phase featuring a reversible -6~eV flat band originating from a $p$-$d$ hybridization collapse. Further strain enhances correlations and reduces the bandwidth of the Cr 3$d$ manifold, producing a correlated Mott-like flat-band regime in which the -6~eV and -2~eV features coexist. Excessive strain causes bond rupture and disorder in the near-surface region, leading to a largely irreversible insulating spectral regime dominated by the -2~eV and -4.5~eV flat bands. (d) Schematic strain-interaction-distortion diagram showing the evolution from an altermagnetic metal through a correlated Mott-like flat-band regime to a disorder-dominated insulating regime with increasing strain.
}
\label{4}
\end{figure*}

By contrast, the deeper -6~eV flat band does not appear in uniform-strain bulk calculations, pointing instead to a strongly strained near-surface strain-gradient environment that lies beyond the reach of conventional DFT. Surface slab calculations with imposed strain gradients would require detailed knowledge of the strained surface geometry and the depth profile of the strain field, which is currently unavailable, making a quantitatively controlled first-principles treatment unfeasible. We therefore rely on experiment to constrain the microscopic picture. The -6~eV band appears only under applied tensile strain and disappears when the strain is released, even though XRD shows that the bulk lattice distortion remains below 1\%, and it is spatially confined to strongly strained regions near the surface. Together, these signatures identify a strong strain gradient localized near the surface as the driving force for this flat band (Fig.~\ref{4}a), and suggest that it corresponds to a deeply bound, structurally localized state formed in a regime of severely stretched but not yet fully ruptured Cr-Sb bonds, distinct from the correlation-induced -2~eV Hubbard-like band.\\

{\bf Discussion and Summary}

The above results suggest a strain-evolution picture in which CrSb is tuned through a sequence of regimes governed by orbital hybridization, electronic correlation and lattice distortion (Fig.~\ref{4}c-\ref{4}d). In the unstrained state, Cr-Cr and Cr-Sb bonds support dispersive Cr~3$d$ bands and a well-defined altermagnetic spin symmetry. Moderate tensile strain reduces the Cr-Sb $p$-$d$ hybridization more rapidly than the Cr-Cr $d$-$d$ channel (Fig.~\ref{4}b), narrowing deep valence bands and generating the structurally driven -6~eV flat band within the near-surface strain-gradient layer. With further strain, the effective Cr-Cr hopping is also suppressed, reducing the Cr~3$d$ bandwidth $W$ and enhancing $U/W$, so that the -2~eV flat band emerges as a predominantly correlation-driven, Mott-like LHB on top of an otherwise intact lattice. Once the system is pushed into the irreversible regime, however, the character of the -2~eV feature changes: a -2~eV component persists but now follows the onset of local structural distortions and bond breaking, rather than purely bandwidth-controlled correlations. We therefore distinguish a correlation-driven, reversible -2~eV Hubbard-like band from a structurally driven, irreversible -2~eV component that reflects disorder-enhanced localization of Cr~3$d$ states, in close analogy to the structurally localized Sb~5$p$ states that form the -4.5~eV band. In this disorder-dominated regime, deep-valence spectral weight associated with the -6~eV features is redistributed into the broad band near -4.5~eV, core levels acquire additional components, and SEM reveals dense stripe-like deformation patterns, together signalling a transition from a hybridization-collapsed but chemically intact state to one with structurally localized electronic states and an insulating spectral response, as summarized in Fig.~\ref{4}c-\ref{4}d.

Since ARPES requires an atomically clean surface, the sample is cleaved in-situ introducing steps and microcracks that generate a strongly non-uniform near-surface stress field under tensile loading. The resulting high-strain surface layer effectively behaves as an elastically strained “film” clamped to a compliant CrSb substrate, capable of sustaining local in-plane distortions without catastrophic fracture, as evidenced by the dramatic Fermi-surface distortions and the emergence of both structural (-6~eV) and correlated (-2~eV) flat bands evolve reversibly within the elastic window. This extreme-strain configuration should therefore be viewed as a high-strain proof-of-principle geometry that demonstrates what becomes possible once such strain gradients are accessed in a crystalline altermagnet: flat-band tendencies emerge within an already spin-textured electronic manifold, bandwidth narrowing and symmetry reshaping become strongly entangled, and an orbital-selective altermagnetic spin texture can persist deep into an insulating spectral regime. In the future, analogous strain-gradient fields could be engineered in thin-film or microstructured altermagnets using patterned stressors, compliant substrates or local actuators, providing more device-oriented platforms that build on the unified strain-correlation-spin-texture control established here. Crucially, in all of these scenarios the correlated flat band would form within a band manifold that already hosts an altermagnetic, orbital-selective spin texture, so strain- and correlation-driven localization becomes spin-, orbital- and momentum-selective in a zero-net-moment background---a combination not realised in spin-degenerate correlated flat-band systems or in weakly correlated altermagnets. This unique intertwining of altermagnetism, extreme strain and flat-band correlations provides a natural basis for spin-selective or orbital-selective Mott filters: the altermagnetic spin splitting separates spin channels in momentum and orbital space, while the strain-narrowed flat band can selectively localise one spin-orbital sector and leave the complementary sector itinerant, enabling highly spin-polarised currents without stray fields. More broadly, the strain-symmetry-correlation design principles established here should be applicable to a wide range of altermagnets and spin-textured correlated materials, also including kagome metals and moir\'e superlattices, providing a general route to co-engineer flat-band tendencies and unconventional spin transport with extreme lattice control.


In summary, the combined ARPES, spatially resolved spectroscopy, XRD, SEM, and DFT results show that tensile strain provides a continuous tuning knob that couples lattice deformation to orbital hybridization, electron correlation, and magnetic symmetry in altermagnetic CrSb. Moderate strain produces two distinct reversible flat bands arising from $p$-$d$ hybridization collapse (-6~eV) and correlation-enhanced localization (-2~eV) within the Cr 3d manifold. Stronger strain induces structural disorder and partial bond decoupling in the near-surface region, leading to a largely irreversible insulating spectral response. This sequence can be summarized in a strain-symmetry-correlation diagram that connects altermagnetic metallic, correlation-enhanced and disorder-dominated insulating regimes. CrSb thus provides a model platform for mechanically tuning quantum correlation and spin texture, and points toward strain-adaptive altermagnetic and correlated spintronic functionalities,  including zero-net-moment spin filters based on spin-selective Mott or orbital-selective Mott localization in correlated flat bands of epitaxially strained CrSb and related altermagnets.\\

\noindent {\bf Methods}\\
\noindent{\bf Sample} The CrSb single crystals were grown by the chemical vapor transport (CVT) method. A stoichiometric ratio of chromium and antimony powders, together with iodine of 2.5 mg/ml as the transport agent, were mixed and sealed in an evacuated quartz ampoule. The ampoule was slowly heated and finally exposed to a temperature gradient of 925°C to 900°C where the CVT preceded for one week, then naturally cooled down to room temperature. CrSb crystals in size of 5-10 mm with regular shapes and shiny surfaces were obtained. 

\noindent{\bf ARPES Measurements} High-resolution ARPES measurements were performed at the Bloch beamline of MAX IV and at the I05 beamline of the Diamond synchrotron light source. The total energy resolution (analyzer and beamline) was set at 15$\sim$20 meV for the measurements. The angular resolution of the analyser was $\sim$0.1 degree. The beamline spot size on the sample was about 10 $\mu$m$\times$10 $\mu$m at the Bloch beamline of MAX IV and about 50 $\mu$m$\times$50 $\mu$m at the I05 beamline of the Diamond synchrotron. The samples were cleaved {\it in situ} and measured at about 18 K at the Bloch beamline of MAX IV and about 8 K at the I05 beamline of the Diamond synchrotron in ultrahigh vacuum with a base pressure better than 1.0$\times$10$^{-10}$ mbar. 

\noindent{\bf Strain device} Our uniaxial strain device is based on a customized bendable sample platform that enables in situ application of large compressive or tensile strain. The design consists of a T-shaped platform, a step-like actuator, a driving screw with a circlip, and a customized Omicron holder. All components are non-magnetic and are typically machined from BeCu, which offers excellent elastic, thermal and electrical properties. As shown in Fig. S2b-S2c in Supplementary Materials, tightening the driving screw pushes the actuator downward and bends the edge-connected platform towards the actuator, thereby imposing tensile strain along the platform; loosening the screw allows the circlip underneath to lift the actuator, bending the platform in the opposite direction and generating compression. In the present geometry this configuration produces strong near-surface strain gradients that are sufficient to induce the pronounced Fermi-surface distortions and flat-band features reported in the main text, while remaining compact and fully compatible with standard spectroscopic manipulators. Within a simple homogeneous-strain approximation, the most strongly deformed regions correspond to an effective in-plane contraction of the surface BZ of order 20\%, as inferred from ARPES (Fig.~\ref{2}), but this value should not be interpreted as a uniform microscopic strain.

\noindent{\bf DFT calculations} 
All band-structure calculations, orbital projections, and partial densities of states were performed using VASP \cite{GKresse_PRB1996_JFurthmuller} with the projector augmented-wave (PAW) method \cite{PEBlochl_PRB1994} and the LDA exchange-correlation functional \cite{LDA_CA}. The Brillouin zone was sampled using a Gamma-centered $10\times10\times8$ $k$-point mesh. The plane-wave energy cutoff was set to 550 eV. A Hubbard U term of 0.8 eV was applied to the Cr d-orbitals within the DFT+U framework unless otherwise noted, in order to account for electron-electron correlations. Strain was applied to the lattice unit cell without further atomic relaxation. All Wannier-based tight-binding Hamiltonians were constructed using the WANNIER90 interface \cite{NMarzai_PRB1997_DVanderbilt}. The projectors included the Cr $d$-orbitals and Sb $p$-orbitals, fitted in the energy window from -2 eV to 2 eV. Fermi-surface calculations were carried out using WannierTools \cite{QSWu_CPC2018_AASoluyanov}.

The dynamics stability for the strained CrSb system were calculated based on the DFT\cite{PHohenberg_PR1964_WKohn,WKohn_PR1965_LJSham} and density functional perturbation theory (DFPT)\cite{FGiustino_RMP2017,SBaroni_RMP2001_PGiannozzi} as implemented in the QUANTUM ESPRESSO (QE) package\cite{PGiannozzi_JPCM2009_RMWentzcovitch}. The generalized gradient approximation (GGA) of Perdew-Burke-Ernzerhof (PBE) type\cite{JPPerdew_PRL1996_MErnzerhof} was chosen for the exchange-correlation functional. The kinetic energy cutoff of the wavefunction was set to be 80 Ry. In the phonon structure calculations, we sampled BZ using a 6$\times$6$\times$6 q-point mesh. The Gaussian smearing method with a width of 0.004 Ry was employed for the Fermi surface broadening. In structural optimization, we employ an orthorhombic 1$\times$$\sqrt{3}$$\times$1 supercell. Where only the lattice of straining direction was fixed, and other lattice constants and internal atomic positions were fully relaxed until the forces on all atoms were smaller than 0.002 Ry/Bohr. 

\noindent {\bf Data Availability}

\noindent The authors declare that all data supporting the findings of this study are available within the paper and its Supplementary Information files.

\vspace{3mm}

\noindent {\bf Acknowledgements}\\
The work presented here was financially supported by the Swedish Research council (2019-00701 and 2019-03486) and the Knut and Alice Wallenberg foundation (2018.0104). We acknowledge MAX IV Laboratory for time on Beamline BLOCH under Proposal 20250284 and 20241395 as well as Diamond Light Source for time on Beamline I05 under Proposal SI39652 and SI41048. Research conducted at MAX IV, a Swedish national user facility, is supported by the Swedish Research council under contract 2018-07152, the Swedish Governmental Agency for Innovation Systems under contract 2018-04969, and Formas under contract 2019-02496.
\vspace{3mm}

\noindent {\bf Author Contributions}\\
Co.L. proposed and conceived the project. Co.L. carried out the ARPES experiments with the assistance from M.H.B., F.S. and O.T.. M.L.H., J.F.Z. and T.X. contributed to the band structure calculations. Z.L.L. contributed to CrSb crystal growth. Co.L. and D.P. contributed to XRD and SEM measurements. Co.L. contributed to software development for data analysis and analyzed the data. Ch.L., W.Y.C. and J.C. supported the strained sample holder. O.C., T.K., J.O., C.P. and B.T. provided the beamline support. Co.L., M.L.H., J.F.Z., M.H.B., T.X. and O.T. contributed to the scientific discussions. Co.L. wrote the paper. All authors participated in and commented on the paper.

\noindent {\bf Competing Interests}\\
The authors declare no competing interests.

\newpage

\end{document}